# IRRADIANCE VARIATIONS DUE TO ORBITAL AND SOLAR INERTIAL MOTION: THE EFFECT ON EARTH'S SURFACE TEMPERATURE

Gerald E. Marsh
Argonne National Laboratory (Retired)
gemarsh@uchicago.edu

## ABSTRACT

Variation in total solar irradiance is thought to have little effect on the Earth's surface temperature because of the thermal time constant—the characteristic response time of the Earth's global surface temperature to changes in forcing. This time constant is large enough to smooth annual variations but not necessarily variations having a longer period such as those due to solar inertial motion; the magnitude of these surface temperature variations is estimated.

# Introduction

Total solar irradiance (TSI) is defined as the irradiance of solar emission at the mean distance of the Earth from the Sun. Since the Earth's distance from the Sun varies as it traverses its orbit, the actual value changes significantly as a function of the day of the year. Measurements of the magnitude of TSI at the mean distance range from about 1360 $w/m^2$ to 1368 $w/m^2$. The value of 1367 $w/m^2$ is quite common and will be used here. Variations in the averaged Solar irradiance is shown in Fig. 1.

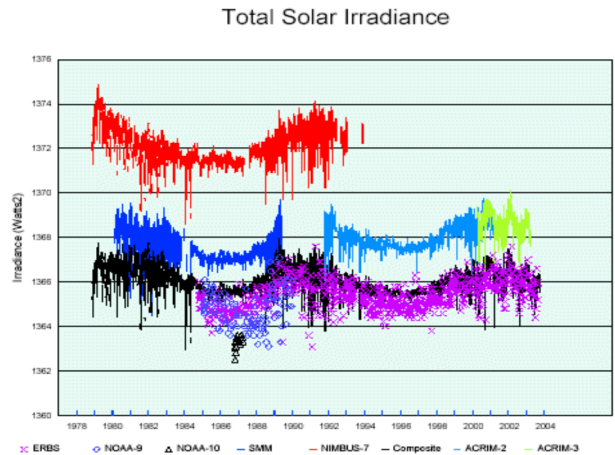

Figure 1. Total Solar Irradiance from six sets of satellite observations. [NOAA National Center for Environmental Information]

The variation in the Earth's orbit around the Sun is shown in Fig. 2. Note that the maximum difference in the distance of the Earth from the Sun is about $5 \times 10^9$ m.

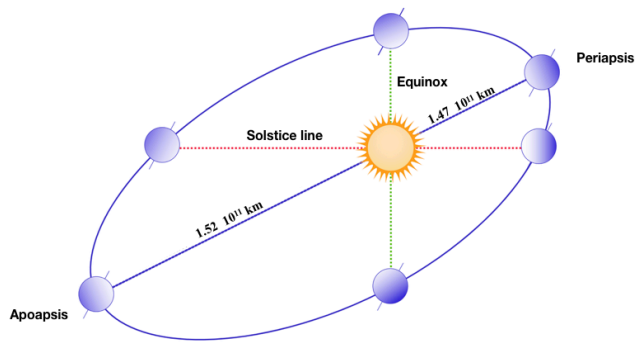

Figure 2. Orbit of the Earth around the Sun. The difference in the apoapsis and periapsis is $5 \times 10^9$ m.

Because the irradiance changes as the square of the distance from the Sun, the variation in solar irradiance is not small. It is given by,

$$I = 1367 \, w/m^2 \left[1 + 0.034 \cos\left(\frac{2\pi n}{365.25}\right)\right],$$

(1)

Where $n$ is the day of the year with January 1st being $n = 1$. This variation is shown in Fig. 3.

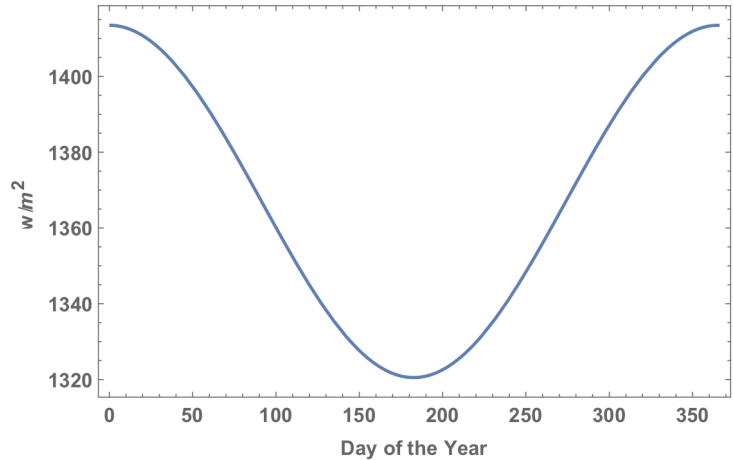

Figure 3. Solar irradiance as a function of the day of the year. The range of the irradiance in traversing the Earth's orbit is 93 $w/m^2$.

These relatively fast variations are thought to average out and not significantly affect the Earth's surface temperature because of the large thermal time constant of the Earth—to be discussed later in this paper.

Another source of solar irradiance variation is the motion of the Sun around the center of mass of the solar system known as the barycenter.[1] This motion is called Solar Inertial Motion (SIM). It is primarily caused by the varying positions of the most massive planets Jupiter, Saturn, Uranus, and Neptune. The motion of the sun takes place within a circular region of diameter of ~4.3 solar radii, or $3 \times 10^9$ m. This should be compared with the maximum difference in the distance of the Earth from the Sun of $5 \times 10^9$ m. An example of such solar inertial motion is shown in Fig. 4.

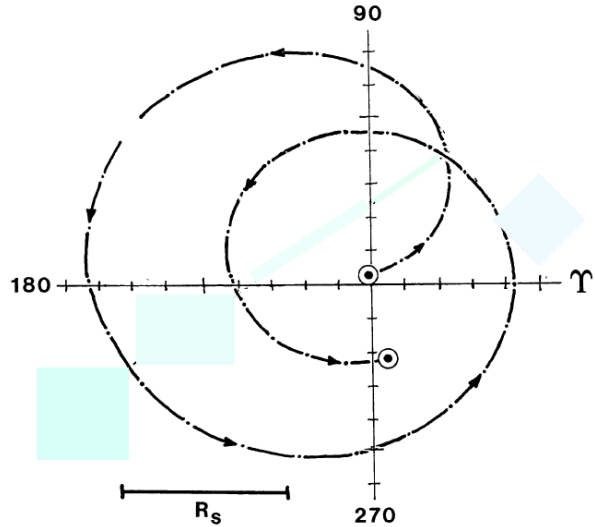

Figure 4. The orbit of the Sun around the barycenter of the solar system during the period 1990 to 2013. The barycenter of the solar system is at the origin and the points on the curve give the position of the center of the Sun at 200-day intervals. The plane of the figure parallels the equatorial plane of the Earth and the vernal equinox, defined as when the plane of the Earth's equator passes through the center of the Sun, is indicated on the abscissa. The units on the axes are AU × $10^{-3}$ or $1.496 \times 10^8$ m. [Adapted from Fig. 1 of R.W. Fairbridge and J. H. Shirley, *Solar Physics* **110** (1987), pp. 191-220.]

Averaging over a variety of such plots, the Sun orbits the barycenter of the solar system about once every 10 yrs. Figure 4 will be used in the analysis that follows primarily because it can be approximated by an analytic expression for a rose thereby greatly simplifying the numerical calculations. In Mathematica an example of a rose is given by ParametricPlot[{(0.4 + 1.6 Cos[$\theta$]) Cos[$\theta$], (0.4 + 1.6 Cos[$\theta$]) Sin[$\theta$]}, {$\theta$, 0, $2\pi$}]. Each loop in Fig.4 also takes roughly 10 yrs (with a velocity varying between 9 *m/s* and 16*m/s*)[2] and this will be assumed to be the case in what follows. A comparison of Fig. 4 and an appropriate rose is given in Fig. 5. Since the purpose here is to estimate the variations in solar irradiance due to solar inertial motion, the small differences between the two curves is of no great importance.

In general, the motion of the Sun around the barycenter of the solar system is more complex than this example with the motion falling into two basic types, which can be characterized as ordered, following trefoils related to the motion of Jupiter and Saturn, and disordered or chaotic.[3] Because motion in the solar system is itself chaotic, the solar inertial motion is intrinsically chaotic.

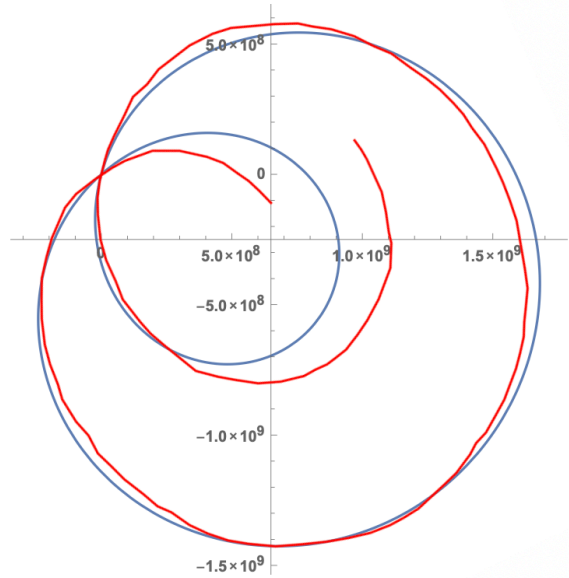

Figure 5. A comparison by superposition of Fig. 4 (red) with an approximation given by a rose (Blue). The greatest discrepancy is where the motion of the Sun does not close.

The relative orientation of this motion with respect to Fig. 2 is shown in Fig. 6. Note that the vernal equinox in Fig. 4 is aligned with the vernal equinox of the solar system.

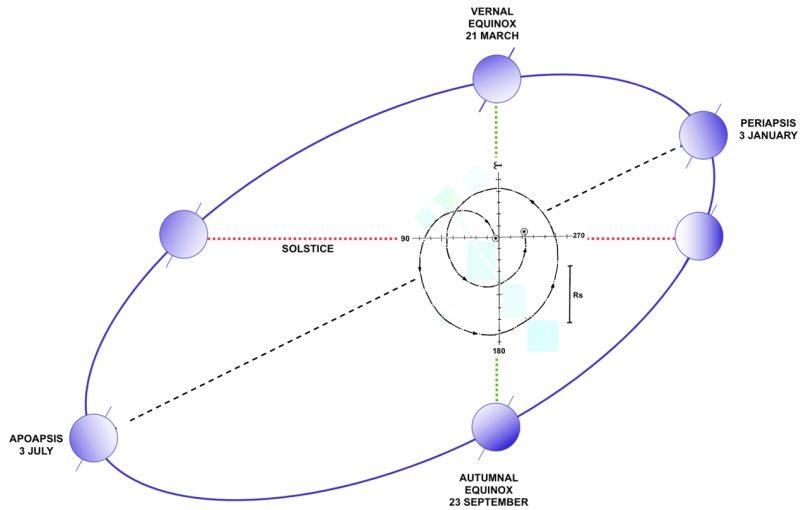

Figure 6. This figure shows the alignment of the vernal equinox in Fig. 4 with that of the solar system. The angles shown in the figure are measured from the vernal equinox line.

**Solar Inertial Motion and Total Solar Irradiance**

The geometry needed to find the variation in TSI is shown in Fig. 7. The distance of the Earth from the barycenter of the solar system, $r_{toE}$, is a function of the angle $\theta$, which itself depends on the position of the Sun on the path given by the rose.

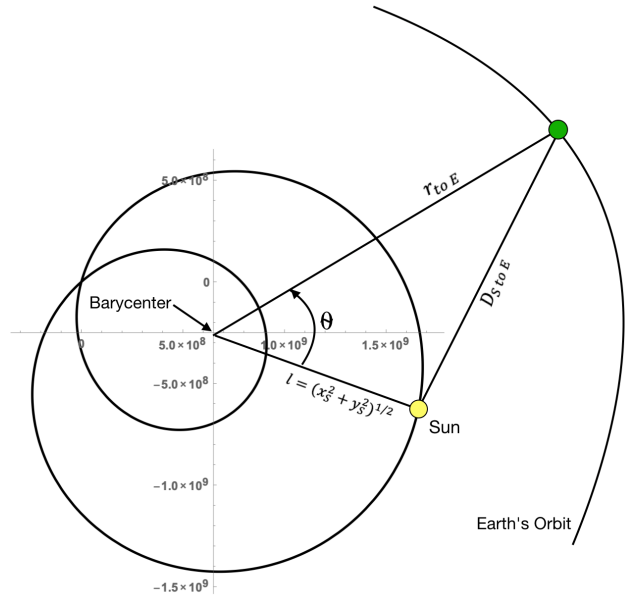

Figure 7. The distance from the Sun to the Earth, $D_{StoE}$, as a function of distance of the Earth from the barycenter of the solar system and the position of the Sun on the rose representing the SIM. The rectangular coordinate system is that associated with the rose and $x_s^2$ and $y_s^2$ are the coordinates of the Sun.

The distance from the Sun to the Earth, $D_{StoE}$, is given by

$$D_{StoE} = (r_{toE}^2 + (x_s^2 + y_s^2) - 2r_{toE}(x_s^2 + y_s^2)^{1/2} \cos\theta)^{1/2}.$$

(2)

The distance from the barycenter to Earth can be found by assuming Earth's orbit is Keplerian, in which case

$$r_{toE} = a \frac{1 - e^2}{1 + e \cos\varphi},$$

(3)

where *a* is the length of the semi-major axis of the Earth's orbit, *e* is the eccentricity of the orbit (currently about 0.0167), and $\varphi$ is the angle subtended at the Sun between the semi-major axis line and the current position of the Earth.

The angle $\varphi$ has the range 0 to 360°. This is very close to our number of days in the year of about 365.2422. If $\varphi$ is in degrees one can use the day of the year for $\varphi$ if one multiplies the day of the year by 360/365.2422 = 0.9856. Because the eccentricity is so small, $1/(1 + e \cos \varphi)$ can be approximated by $(1 - e \cos \varphi)$. To express the result in AU, $a = 1$ and in addition $(1 - e^2)$ can be set equal to unity. This gives for Eq. 3

$$r_{toE} = 1 - 0.0167 \cos(0.9856 \, Day).$$

(4)

The *Day* is the number of days since the crossing of the perihelion. The Earth currently crosses the perihelion on about January 4th. So one can set *Day* = (*n* − 4), where *n* = 4,5, . . . , 365. To now put the cosine argument in terms of radians one must multiply by $1.745 \times 10^{-2}$; and to convert the day number into radians one must also multiply by $365/2\pi = 58.09$, so that the final approximate expression including these conversions is

$$r_{toE} = 1 - 0.0167 \cos[(1.72 \times 10^{-2})(58.1\varphi - 4),$$

(5)

where $0 \leq \varphi \leq 2\pi$. A plot of $r_{toE}$ as a function of $\varphi$ is shown in Fig. 8

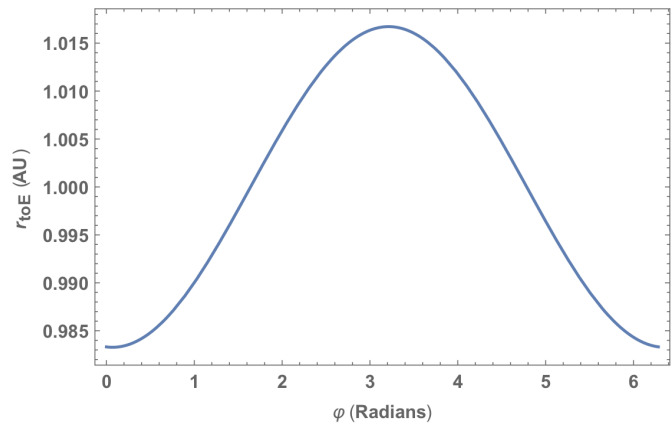

Figure 8. Variation of the distance from the barycenter to the Earth as a function of $\varphi$.

As mentioned earlier, it will be assumed that a circuit around the barycenter of the solar system by the Sun takes roughly 10 yrs so that Eq. (2) yields the rather complicated trigonometric expression

$$
\begin{aligned}
D_{StoE} = \surd(&\left(4.\times 10^8 + 1.6\times 10^9 \text{Cos}\left[0.558 + \frac{\theta}{10}\right]\right)^2 \text{Cos}[\theta]^2 \\
&+ 2.238016\times 10^{22}(1 - 0.01672\text{Cos}[0.01719872(-4 + 58.09\theta)])^2 \\
&+ \left(4.\times 10^8 + 1.4\times 10^9 \text{Cos}\left[0.558 + \frac{\theta}{10}\right]\right)^2 \text{Sin}[\theta]^2 - 2.992\times 10^{11}\text{Cos}[\theta](1 \\
&- 0.01672\text{Cos}[0.01719872(-4 \\
&+ 58.09\theta)])\surd((4.\times 10^8 + 1.6\times 10^9 \text{Cos}[0.558 + \frac{\theta}{10}])^2 \text{Cos}[\theta]^2 \\
&+ (4.\times 10^8 + 1.4\times 10^9 \text{Cos}[0.558 + \frac{\theta}{10}])^2 \text{Sin}[\theta]^2))
\end{aligned}
$$
(6)

where distances are now in meters.

The solar luminosity is $3.8\times 10^{26}$ watts so that the solar irradiance is given by SolIrrad = $3.8\times 10^{26}/4\pi D_{StoE}^2$. The plot of SolIrrad is shown in Fig. 9. The high frequency component is from the Earth's motion around the Sun (see Fig. 3), which is modulated by the solar inertial motion. The magnitude of the latter can be more easily seen from the upper envelope of the plot as seen in Fig. 9.

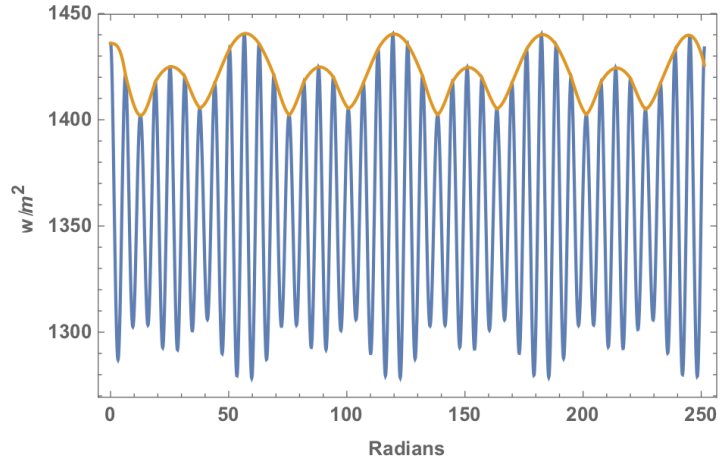

Figure 9. Solar irradiation along the Earth's orbit modulated by the influence of SIM. The abscissa has the range $0 \leq \theta \leq 80\pi$ corresponding to 40 yrs, or two full cycle around the rose. The upper envelope of SolIrrad is also shown. The high frequency component comes from the yearly period of the Earth around the Sun. The maximum peak-to-peak magnitude due to the SIM is about 38 w/m².

The variation in solar irradiance due to solar inertial motion is then about 2.7%. The real question is how this affects the surface temperature of the Earth. The answer depends on the Earth's heat capacity and thermal time constant.

**Heat Capacity and Time Constant of the Earth**

A good deal of research has been done to estimate the heat capacity and time constant of the Earth. Schwartz[4] found that the equilibrium sensitivity of the Earth's climate is determined by the quotient of the relaxation time constant and the effective heat capacity of the climate system. That is,

$$\lambda_s^{-1} = \tau/C,$$

(7)

where $\lambda_s^{-1}$ is the equilibrium sensitivity (note that the exponent is included in the definition). The effective heat capacity, $C$, is that portion of the global heat capacity that couples to a perturbation of the climate system for the length of time of the perturbation. Schwartz estimated this heat capacity and found that if the planetary coalbedo—defined as the fraction of incoming solar energy that is absorbed by the climate system—is held constant, the time constant of the global climate system ~5 years. Serious questions have been raised by Knutti, et al.[5] in responded to this paper saying that the implication of such a short time constant is that the global surface temperature is nearly in equilibrium with radiation forcing. Based on nineteen 3-dimensional coupled atmosphere-ocean general circulation models they argue that the time scale should be larger by about at least a factor of three. In addition, they find that the linear relation of Eq. (7) between climate sensitivity and the time factor is unwarranted. They both essentially agree on Schwartz's estimate of the effective thermal heat capacity being about $17 \pm$ *W yr $m^{-2}K^{-1}$*. We will return to the issue of the Earth's time constant and how it affects the surface temperature of the Earth later in this paper.

In order to estimate the change in the Earth's surface temperature use will be made of the one-dimensional radiative-convective equilibrium model of Hartmann.[6] The model is based on an average cloud cover and includes water-vapor feedback. The response of the model to changes in TSI and changes in carbon dioxide concentration is shown in Fig. 10.

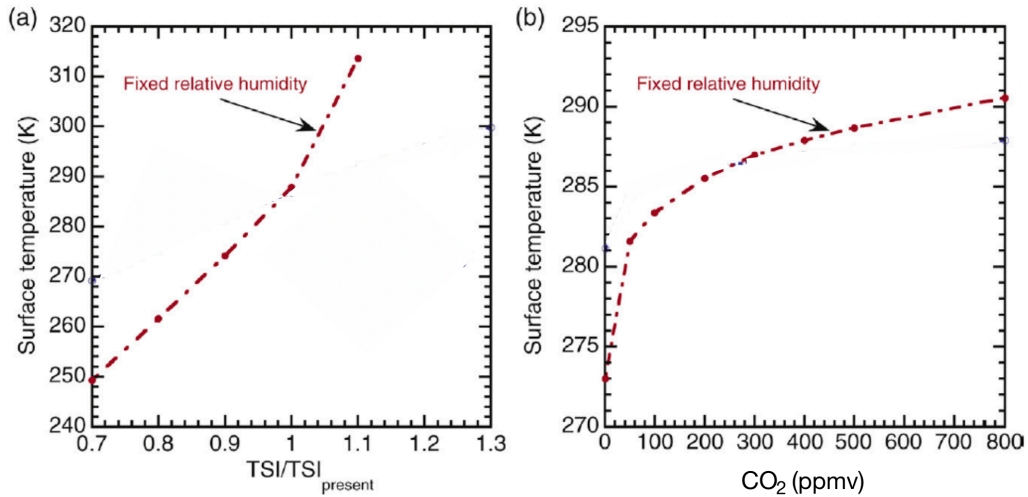

Figure 10. The one-dimensional radiative-convective equilibrium model of Hartmann. The figure is adapted from Hartmann's Fig. 10.2.

The model shows that the change is surface temperature due to an increase in $CO_2$ concentration from its current value of just over 407 ppmv to 800 ppmv would change the surface temperature from 288 °K to about 290 °K or 2 °C consistent with estimates by the IPCC. Using this model leads to the equilibrium changes in surface temperature shown in Fig. 11.

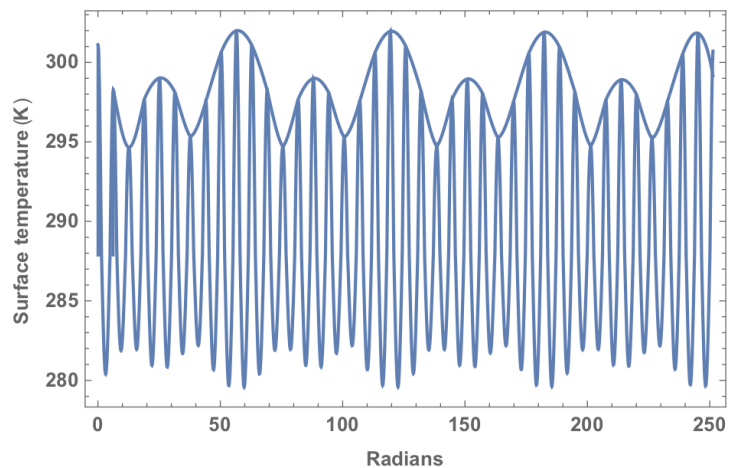

Figure 11. Equilibrium change in surface temperature due to changes in TSI induced by SIM. The maximum change in surface temperature peak-to-peak estimated from the upper envelope is 7.4 °C, giving an approximate amplitude for the trigonometric curve of ~3.7 °C.

While Figure 11 shows the equilibrium change in surface temperature, it does not take into account the Earth's time constant. Pierrehumbert[7] addresses the issue of the thermal inertia of the Earth by using a differential equation similar to that for a series resistance and capacitance electrical circuit driven by a time varying voltage. He uses an example where the change in solar radiation is given by a cosine function, and finds that the amplitude of the temperature variation $A$ is given by

$$|A| = Const. \frac{1}{\sqrt{1 + (\omega\tau)^2}},$$

(8)

where the constant is the amplitude found when $\omega\tau \ll 1$.

Here, the change in solar irradiance, although it is a trigonometric function, it is not a simple one (see Eq. (6) and the definition of SolIrrad that follows). The relevant differential equation is linear and of the first order and is of the form

$$\frac{dT(t)}{dt} + \frac{T(t)}{\tau} - \frac{SolIrrad}{\tau} = 0.$$

(9)

The non-transient part of the solution is

$$T(t) = e^{-t/\tau} \int_1^t \frac{e^{-t/\tau} SolIrrad}{\tau} dt,$$

(10)

but, unfortunately, the integral is intractable. Nonetheless, if one applies the factor $1/\sqrt{1 + (\omega\tau)^2}$ to the 3.7 °C variation due to SIM in Fig. 11, the results for the two time constants $\tau = 5yr$ and $\tau = 10yr$ discussed above—with $\omega = 0.63$ from the 10 yr period of the SIM are

| $\tau$ (yr) | $\omega\tau$ | $1/\sqrt{1+(\omega\tau)^2}$ | $\Delta T_s$ (°C) |
|---|---|---|---|
| 5 | 3.15 | 0.302 | 1.2 |
| 10 | 6.3 | 0.156 | 0.58 |

Table 1. Effect of the factor $1/\sqrt{1+(\omega\tau)^2}$ on the variation of $T_s$ due to SIM.

For a one-year period and $\tau$ = 5yr and $\tau$ = 10yr, $\Delta T_s$ is 0.12 °C and 0.06 °C respectively. One can see that the longer the time constant the smaller the change in surface temperature $\Delta T_s$. The SIM effect on $T_s$ effectively vanishes for $\tau \geq 15 yr$.

It is possible to get an empirical estimate of $\tau$. Figure 12 shows the global average cloud cover and temperature anomaly from 1983 to 2008. These are obviously closely correlated and can be used to estimate the range of $\tau$. A possible explanation for the drop in cloud cover is given in Appendix 1.

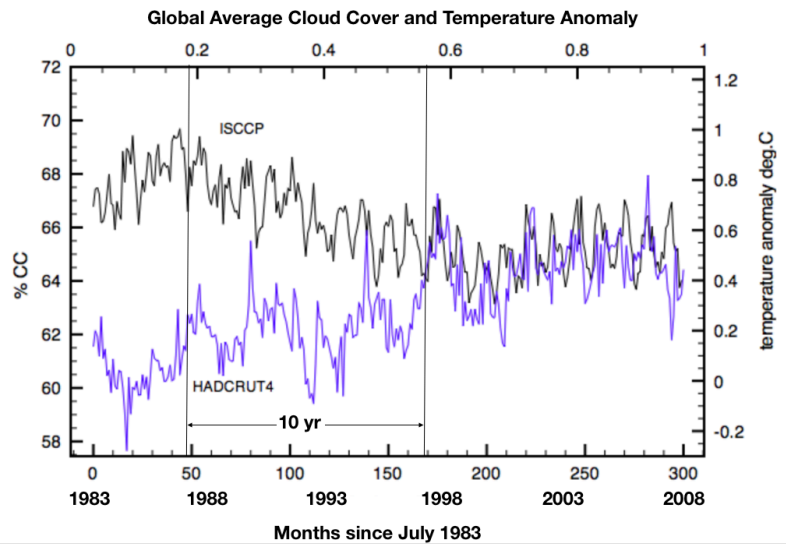

Figure 12. Global average monthly cloud cover from the International Satellite Cloud Climatology Project (red) and monthly surface temperature anomalies from Met Office Hadley Centre and the Climatic Research Unit at the University of East Anglia (HADCRUT4) in blue. The reference period for HADCRUT4 is 1961-1990.

Hartmann has estimated that a 10% change in cloud cover corresponds to a doubling of carbon dioxide concentration, which itself corresponds to ~2 °C temperature rise [See Fig. 10(b)]. Thus the 2% drop in cloud cover from 1987-1997 shown in Fig. 12 corresponds to essentially all of the temperature rise of ~0.3 °C during these years. Note that from 1998 to 2008 there was very little change in cloud cover and temperature, which continue to track each other. The ten years from 1987-1997 show that the increase in insolation is closely followed by the rise in temperature although the rate of rise is slower until just before 1997. The implication is that the Earth's time constant could well lie in the range of $5yr \leq \tau \leq 10yr$, and is probably closer to 10yr.

**Summary**

The effect on total solar irradiance due to solar inertial motion was found by using an approximation of a geometrical heart to represent the motion of the Sun around the barycenter of the solar system for a representative SIM orbit; this was shown in Fig. 9. The upper envelope of the resulting complex trigonometric function was then used in conjunction with the equilibrium one-dimensional radiative-convection model of Hartmann to convert the data in Fig. 9 to surface temperature variations as shown in Fig. 11. The thermal inertia of the Earth, as captured in its time constant, was discussed and the effect of this time constant on the surface temperature variations was estimated using an empirical approach based on the correlation between the change in global cloud cover and temperature. The result was a time constant close to 10yr; this is closer to the time constant estimated by Knutti, et al. rather than that given by Schwartz. The conclusion is that variations in Earth's surface temperature due to SIM is comparable to that due to the changes in global cloud cover and temperature change during the period of 1987-1997. Such variations should be taken into account when estimating surface temperature changes of the Earth due to other types of forcing.

# Appendix 1

Figure 12 raises the question of what could be the reason for the drop in cloud cover in the interval indicated. It is not necessary to understand the reason for this drop for the purposes of this paper, but it is an interesting question. Clouds are the *bête noir* of climate modeling but there is some work that has been done on cloud formation to explain variations in low cloud cover. Figure 13 is an example of the result of this research from almost twenty years ago.

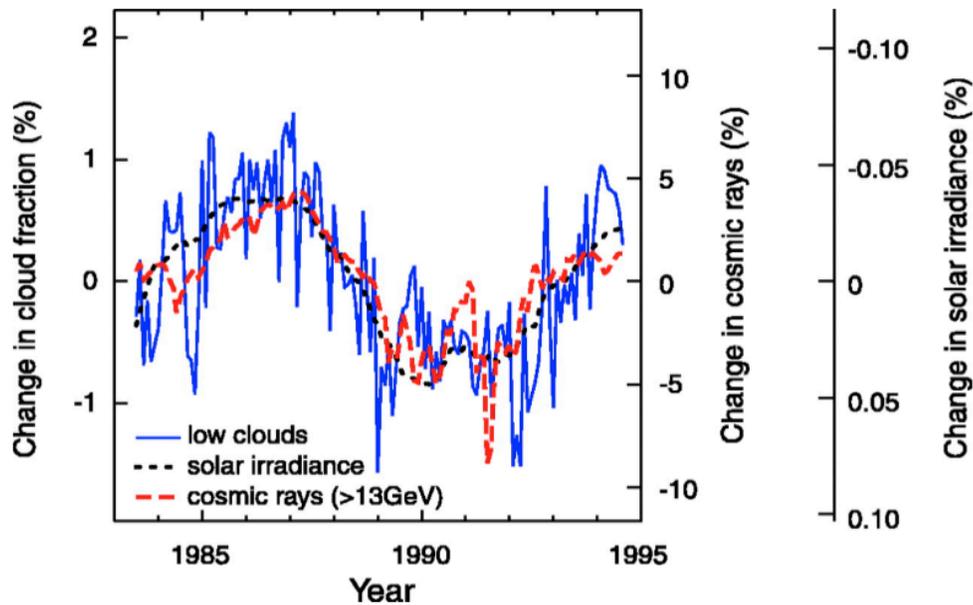

Figure 13. Variations of low-altitude cloud cover (less than about 3 km), galactic cosmic rays, and total solar irradiance between 1984 and 1994. From K.S. Carslaw, R.G. Harrison, and J. Kirkby, *Science* **298**, 1732 (2002). Note the inverted scale for solar irradiance.

What Fig. 13 shows is the very strong correlation between galactic cosmic rays—those having an energy greater than 13GeV, solar irradiance, and low cloud cover. Varying solar activity is not only associated with changes in irradiance, but also with changes in the solar wind, which in turn affect cloud cover by modulating the cosmic ray flux. This, it is argued, constitutes the strong positive feedback needed to explain the significant impact of small changes in solar activity on climate.

The key issue is whether or not ionization due to these galactic cosmic rays can significantly increase the number of cloud condensation nuclei or the rate of their formation. There is an ongoing experiment at CERN called the CLOUD experiment. From initial results, Kirkby, et al. found that, "Ions increase the nucleation rate by an additional factor of between two and more than ten at ground-level galactic-cosmic-ray intensities, provided that the nucleation rate lies below the limiting ion-pair production rate."[8] However, the effect was judged to be small. As of 2019, CERN has started an experiment of the "CLOUDy" type with a generator of electrically charged cloud seeds to investigate the effects of charged aerosols on cloud formation. The issue of the magnitude of the relation between galactic cosmic rays and cloud formation and rate is still open.